\documentclass[epj]{svjour}
\usepackage{graphicx,times}
\usepackage[T1]{fontenc}
\newcommand{\pslash}{p\hspace{-.5em}/\hspace{.15em}}
\newcommand{\ket}[1]{\left| #1 \right>}
\newcommand{\qqbar}{{q\overline{q}}}
\newcommand{\uubar}{{u\overline{u}}}
\newcommand{\ddbar}{{d\overline{d}}}
\newcommand{\ssbar}{{s\overline{s}}}

\begin{document}

\title{$U_A(1)$ anomaly and $\eta^\prime$ mass from an infrared singular 
quark-gluon vertex}
\author{Reinhard Alkofer\inst{1}, Christian S.\ Fischer\inst{2}\inst{3} and Richard
 Williams\inst{2}}
 \authorrunning{Alkofer \emph{et al}}
\institute{Graz University, Universit\"atsplatz 5, A-8010 Graz, Austria 
      \and Institut f\"ur Kernphysik, TU Darmstadt, Schlossgartenstrasse 9, 64289 Darmstadt, Germany 
	\and Gesellschaft f\"ur Schwerionenforschung mbH, Planckstr. 1  D-64291 Darmstadt, Germany
}
\date{September 8, 2008}
\def\makeheadbox{} 

\abstract{
The $U_A(1)$ problem of QCD is inevitably tied to the infrared behaviour of
quarks and gluons with its most visible effect being the  $\eta^\prime$ mass. 
A dimensional argument of Kogut and Susskind showed that the mixing of the 
pseudoscalar flavour-singlet mesons with gluons can provide a screening of the
Goldstone pole in this channel if the full quark-quark interaction is strongly
infrared singular as  $\sim 1/k^4$. We investigate this idea using previously
obtained results for
the Landau gauge ghost and gluon propagator, together with
recent determinations for the singular behaviour of the quark-gluon vertex.
We find that, even with an infrared vanishing gluon propagator, the singular
structure of the quark-gluon vertex for certain kinematics is apposite for
yielding a  non-zero screening mass.
} 
\maketitle

\section{Generalities on the $U_A(1)$ Anomaly\label{Sec:Intro}}
It has now been long accepted that the strong interaction is described
by a Yang-Mills theory, whose non-Abelian character exhibits the properties
of asymptotic freedom and confinement. However, despite our accedence to
this we still have no satisfactory understanding of the confinement
phenomena itself. It is
envisaged that it is in the infrared structure of QCD where this
mechanism lies, and so we must probe in a momentum region characterised by
large values of the coupling. Here, standard tools such as perturbation
theory are inapplicable and lattice calculations 
typically 
use volumes too small to reliably probe the deep infrared.

The idea of `infrared slavery' gave rise to two possible behaviours 
for the strong running coupling at small momenta \cite{Weinberg:1973un}:
either it exhibits an IR fixed point or diverges at the origin. 
From functional methods in Landau gauge
it is the former behaviour that seems likely in the Yang-Mills
sector~\cite{von Smekal:1997is,Lerche:2002ep,Pawlowski:2003hq,Alkofer:2004it,Fischer:2006vf}. 
However, such a fixed point behaviour of the coupling 
does not predicate that the vertex functions themselves are
infrared finite. Indeed, depending upon the number of external ghost and
gluon legs, many 1PI Green's functions are found to be singular in the
infrared~\cite{Alkofer:2004it,Fischer:2006vf,Alkofer:2008jy}. For the
quark-gluon vertex, this issue is addressed in~\cite{Alkofer:2008tt}.
In this paper it is our aim to address yet another puzzle of QCD, with the
infrared behaviour of the quark-gluon vertex playing a crucial r\^ole.

We know that the up and down quark are very light, and so may consider
them massless to good approximation. To a lesser extent, the same may be
assumed for the strange quark and so we expect the Lagrangian of QCD to
exhibit an approximate $SU(N_f)\times SU(N_f)$ symmetry, with $N_f=3$.  
Assuming the chiral limit is realised, this is broken dynamically to a 
diagonal $SU_V(3)$, and so gives rise to eight Goldstone bosons. 

However, we also expect the same mechanism to break the $U_A(1)$
symmetry of the QCD Lagrangian. Thus one should find a ninth isosinglet
pseudoscalar boson corresponding to the spontaneous breaking of this
symmetry. However, on examining the spectrum of observed mesons this
proves to be one Nambu-Goldstone boson too many. The lightest candidate for this
is the $\eta'$, whose mass of $m_{\eta^\prime}\sim 958$ MeV is far from being `light'.
In fact, it has been shown that if the $U_A(1)$ symmetry is not
explicitly broken, then the mass of the $\eta'$ must be less than
$\sqrt{3}m_\pi\sim 250$ MeV~\cite{Weinberg:1975ui}. Even allowing for the mass of
the strange quark, the $\eta'$ remains far too heavy to be solely described in
this manner. This is the $U(1)$ problem of QCD. 

A step towards resolution came from the recognition that 
the classical $U_A(1)$ symmetry is anomalous, \emph{i.e.} broken by quantum
mechanical effects. However, it is found that the associated current is 
a total divergence, and hence
no symmetry breaking contribution is obtained for \emph{finite} order in
perturbation theory~\cite{Fritzsch:1973pi}. Thus, the mechanism of the
anomalous breaking of the $U_A(1)$ symmetry must be wholly
non-perturbative in nature.

There have been several suggestions as to how the $\eta'$ obtains its unexpectedly 
large mass. In the early days of QCD Kogut and Susskind pointed out, on dimensional grounds, 
that a contribution to the $\eta'$ in the chiral limit could be obtained by the 
mixing between two infrared enhanced gluons, with momentum space propagator 
$D(k)\sim 1/k^4$ for $k^2\rightarrow0$~\cite{Kogut:1973ab}. A few years later 
topological solutions termed \emph{instantons} were discovered in QCD~\cite{Belavin:1975fg}. 
In contradistinction to the confinement driven Kogut-Susskind scenario, these gave
rise to an alternative solution of the $U_A(1)$ problem: Instantons were shown 
by 't Hooft to lead to the non-conservation of the axial charge and so induce 
a $2N_f$ fermion operator giving rise to a non-zero $\eta'$-mass in the chiral  
limit~\cite{'t Hooft:1976up,'t Hooft:1986nc}. Again a few years later Witten and 
Veneziano \cite{Witten:1979vv,Veneziano:1979ec,Veneziano:1980xs}
proposed their solution of the problem by considering an expansion of QCD
in $N_f/N_c$, where $N_f$ and $N_c$ are the number of flavours and colours respectively.
They showed that the correct pattern of the $U_A(1)$-symmetry breaking can be obtained
if the anomalous mass of the $\eta'$ is related to the topological susceptibility of the 
theory. Since this susceptibility is not necessarily generated by instantons Veneziano
named his paper `$U(1)$ without instantons' \cite{Veneziano:1979ec}.
Indeed, a different type of topologically non-trivial gluon field
configurations leading to a non-vanishing topological susceptibility are center 
vortices as \emph{e.g.} discussed in ref.\ \cite{Engelhardt:2000wc}.

Certainly, with the Witten-Veneziano mechanism at hand confinement and topology based 
explanations for the $U_A(1)$ anomaly are not mutually exclusive. The reason is simply that
well-established suggestions for topology driven mechanisms for confinement exist
\cite{Greensite:2007zz}. Thus the same topologically non-trivial gauge field configurations
could be responsible for both confinement and the resolution of the $U_A(1)$-problem.

In this letter we concern ourselves with the Kogut-Susskind mechanism employing an 
approach similar to that given in~\cite{von Smekal:1997dq,mecke:thesis}. We employ 
solutions to the Dyson-Schwinger equations for the propagators of QCD and the 
quark-gluon vertex to determine the anomalous mass of the $\eta'$ in the chiral 
limit. The technical details are discussed in sections \ref{Sec:fmix} - \ref{Sec:inputs}. 
We present our result in section \ref{Sec:results} and discuss its relation to 
the Witten-Veneziano mechanism in section \ref{topo}. We summarise in section 
\ref{Sec:sum}. 
We wish to note that all calculations are done in Euclidean momentum space.
Furthermore, we will ignore all effects from isospin breaking and work in the
isospin limit.

\section{Flavour mixing\label{Sec:fmix}}

A suitable basis to describe the flavour content of mesons is given by 
the $SU(3)$ singlet and octet basis:
\begin{eqnarray}
\ket{\pi^0} &=&  \left(
\ket{\uubar}-\ket{\ddbar}\right)/\sqrt{2}\nonumber\\
\ket{\eta_8} &=&  \left(
\ket{\uubar}+\ket{\ddbar}-2\ket{\ssbar}\right)/\sqrt{6}\label{eqn:so-basis}\\
\ket{\eta_0} &=&  \left(
\ket{\uubar}+\ket{\ddbar}+\ket{\ssbar}\right)/\sqrt{3}\nonumber
\end{eqnarray}
The octet-singlet mass squared matrix in the isospin limit is thus given by
\begin{eqnarray}
M^2=
 \left(
 \begin{array}{ccc}
  M_\pi^2 & 0        & 0               \\ 
  &&\\
  0       & M_{88}^2 & M_{80}^2 	   \\
  &&\\
  0       & M_{08}^2 & M_{00}^2 +m_A^2\\
 \end{array}
 \right) \;,
 \label{eqn:massmatrixelements2}
\end{eqnarray}
with matrix elements
\begin{equation}
\begin{array}{ccccccc}
&&M_{88}^2 &=&  \frac{2}{3}\left( m_{\ssbar}^2+\frac{1}{2}m_\pi^2\right)
         &=& \frac{1}{3}\left( 4m_K^2-m_\pi^2\right)\nonumber\\
	   &&&&&&\\
M_{80}^2 &=&  M_{08}^2 &=& \frac{\sqrt{2}}{3}\left( m_\pi^2-m_{\ssbar}^2\right)
         &=& \frac{2\sqrt{2}}{3}\left( m_\pi^2-m_K^2 \right) \nonumber \\
	   &&&&&&\\
&&M_{00}^2 &=&  \frac{2}{3}\left( \frac{1}{2}m_{\ssbar}^2+m_\pi^2\right)
         &=&  \frac{1}{3}\left( 2m_K^2+m_\pi^2\right).  	        \nonumber
\end{array}
\end{equation}
Here, we have employed  $m_{\ssbar}=2m_K^2-m_\pi^2$ from the 
Gell-Mann-Oakes-Renner relation to make the
substitution on the right-hand side.  With the basis of
(\ref{eqn:so-basis}) the pion 
is decoupled in the isospin limit. Thus we concentrate on the
$2\times2$ sub-matrix that mixes the $\eta$ and 
$\eta'$ \cite{Veneziano:1979ec}:
\begin{equation}
\frac{1}{3} \left(
 \begin{array}{cc}
  4m_K^2-m_\pi^2 & 2\sqrt{2}(m_\pi^2-m_K^2)\\
   & \\
  2\sqrt{2}(m_\pi^2-m_K^2) &\;\;2m_K^2+m_\pi^2 + 3 m_A^2
 \end{array}
 \right)\;.
\end{equation}

The anomalous mass term, $m_A^2$ is related to the topological
susceptibility $\chi^2$ through the Witten-Veneziano formula  
\cite{Witten:1979vv,Veneziano:1979ec,Veneziano:1980xs}
\begin{equation}
m_A^2 = 2\frac{N_f}{f_0^2}\chi^2\,,  \label{WV}
\end{equation} 
which includes the pion decay constant $f_0 \simeq 93$MeV. By now its derivation 
has been refined and it has been adapted to lattice gauge theory and found 
to nicely agree with theoretical expectations and experimental data
\cite{Seiler:1987ig,Giusti:2001xh,Giusti:2004qd,Luscher:2004fu,DelDebbio:2004ns},
with values reported of the order
$\chi^2\sim\left( 191\pm 5\,\textrm{MeV}\right)^4$.
It is important to note, however, that the corresponding 
Witten-Veneziano mechanism
does not make any reference as to which degrees of freedom are
responsible for the generation of $\chi^2$. We come back to this
point in section \ref{topo}.

The $\eta_0$ and $\eta_8$ are not the physical mass eigenstates, 
and so we diagonalise the $2\times2$ mass-matrix by introducing the
following unitary transformation:
\begin{eqnarray}
 (\eta\ \ \ \eta') &=&
 \left(
 \begin{array}{cc}
  \cos\theta & -\sin\theta\\
  \sin\theta & \cos\theta
 \end{array}
 \right)
 \left( \begin{array}{c} \eta_8\\
				\eta_0
 \end{array}\right)\;.
\label{eqn:etamixing}
\end{eqnarray}
with $\theta$ the singlet-octet mixing angle. This gives rise to the following mass eigenstates:
\begin{equation}
m_\pm^2 =  m_K^2 +\frac{m_A^2}{2}\pm \Delta\;,
\end{equation}
with $m_+$, $m_-$ corresponding to the $\eta'$ and $\eta$ respectively,
and the shift $\Delta$ given by:
\begin{equation}
	\Delta=\sqrt{\left( m_K^2-m_\pi^2
\right)^2-\frac{m_A^2}{3}\left( m_K^2-m_\pi^2
\right)+\frac{m_A^4}{4}}\;.
\label{eqn:massmatrixeigenvalues}
\end{equation}
A phenomenological value for the mixing angle can be calculated from:
\begin{equation}
\tan\theta=-\frac{1}{2\sqrt{2}}\left[ 1-\frac{3}{m_K^2-m_\pi^2}\left(
\frac{m_A^2}{2}-\Delta \right) \right]\;,
\label{eqn:calcmixingangle}
\end{equation}
with the current values $m_K=498$ MeV, $m_\pi=135$ MeV~\cite{Yao:2006px}.
Here, the required input is the anomalous mass contribution $m_A^2$. The
Random Instanton Liquid Model provides a mixing angle of
$\theta\simeq-11.5^\circ$~\cite{Shifman:1978zp,Alkofer:1989uj},
whilst knowledge of the $\eta$ and $\eta'$
masses, respectively about $547$ and $958$MeV, suggests $\sim-20^\circ$.
A detailed study of the available observables yields an average 
value of $-15.4^\circ$~\cite{Feldmann:1998vh}.

\section{Anomalous Contribution of the Gluon\label{Sec:anomgluon}}
If we ignore the contribution from the gluon anomaly, that is by setting
$m_A^2=0$, we would expect
the flavour content of our $\eta$, $\eta'$ to be analogous to that of
the $\omega$, $\rho$ system where ideal mixing is realised to very good
approximation. Diagonalising (\ref{eqn:massmatrixelements2}) entails that the 
mass of the eta be degenerate with that of the pion, with flavour content 
$\left( \uubar+\ddbar \right)/\sqrt{2}$, whilst the $\eta'$ would be a pure 
$\ssbar$ pseudoscalar. We know this to be far from the truth since this
would provide a mixing of $\theta\simeq-54.7^\circ$, hence the 
$U_A(1)$-problem and the desire
to include anomalous contributions from the gluon.

To do so in the Dyson-Schwinger and Bethe-Salpeter approach is a
formidable task, for presently we are limited by kernels that include
only re-summations of gluon ladders. These do not include the necessary
annihilation (and consequent recombination) of $\qqbar$ pseudoscalar
meson into two gluons. Typically, the $U_A(1)$-breaking term is put in by
hand and adjusted as a free parameter to obtain phenomenological
results.

\begin{figure}[h]
\includegraphics*[width=0.99\columnwidth]{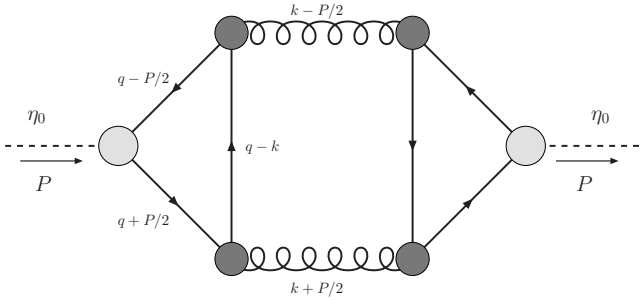}
\caption{The diamond diagram $\Pi(P^2)$. The crossed gluon exchange
contribution gives rise to an additional factor of two.}
\label{fig:diamond}
\end{figure}

The minimal diagram that includes the gluon Adler-Bell-Jackiw (ABJ) 
diagram is given in 
fig.~\ref{fig:diamond}. We later comment upon the contribution of
diagrams with more than two gluon exchange.

\noindent For a simple model, Kogut and Susskind gave arguments as to
how a non-vanishing contribution to the 
$\eta'$-mass in the chiral limit 
could be obtained~\cite{Kogut:1973ab}. 
This has been explored
phenomenologically in a previous study~\cite{von
Smekal:1997dq,mecke:thesis}, in
order to gauge the magnitude of the effect should it be realised.
There, an ansatz for the gluon propagator has been employed which 
is infrared singular supplemented by a perturbative tail.  
Though such a form cannot naturally arise from
consideration of gluons alone, at the time there were already suggestions 
that in Landau gauge the ghost sector was strongly infrared enhanced and
could thus enter into the effective gluon propagator implicitly. 
Using parameters given in that paper, a screening-mass squared of about half the
required amount is obtained, giving credence to the mechanism. However, nothing
definitive may be claimed until dressed Green's functions are employed,
whose behaviour is well-founded in the infrared.

In order to move beyond this qualitative study, we use explicit
calculations for the gluon obtained in previous
calculations~\cite{Fischer:2003rp,Alkofer:2003jj}.
Since these prove to be infrared vanishing, it is important that we
have a self-consistent and motivated dressing for the quark-gluon vertex
that compensates for this precisely, again leading to a non-vanishing
topological susceptibility in the chiral limit.  Our starting point for the 
calculation is the diagram shown in
fig.~\ref{fig:diamond}. It consists of two anomalous AVV triangle
diagrams~\cite{Adler:1969gk}, which here are obviously
non-Abelian in nature~\cite{'t Hooft:1976up}. These triangles are linked
via the exchange of two gluons. 
Note that should we exchange three
gluons or more in the $t-$channel, one would couple in a pseudoscalar $J^{PC}=0^{-+}$
glueball, whose large mass of $\sim2$ GeV~\cite{Narison:1996fm} would heavily suppress the
contribution. For multiple gluon exchange in the $s-$channel, however we
essentially re-sum a gluon ladder which is related to meson
exchange~\cite{Lakhina:2007vm}. For this, the scalar part of the vertex
which we have neglected to model here is important.

Staying with two-gluon exchange and thus restricting ourselves to a
qualitative study, we
write the
amplitude of fig.~\ref{fig:diamond} explicitly as an integral over the gluon momenta:
\begin{equation}
\Pi(P^2) = \int \widetilde{dk}
G^{ac}_{\mu\rho}(P,k)
D^{\mu\nu}_{ab}(k_+)D^{\rho\sigma}_{dc}(k_-)G^{db}_{\sigma\nu}(-P,k)\;,
\label{eqn:amplitude}
\end{equation}
with $\widetilde{dk}$ shorthand for the invariant measure
$d^4k/(2\pi)^4$, and $k_\pm=k\pm P/2$. We have chosen an equal momentum
partition to simplify the resulting equations; the final result should
be independent of this choice. The quark triangle, $G_{\mu\nu}^{ab}$,
arising from the axial anomaly is:
\begin{eqnarray}
G_{\mu\nu}^{ab}(P,k) &=&  g^2\int \widetilde{dq}
\Bigg\{ 
	\textrm{tr}\bigg[
	\Gamma_\eta(P,q)S(q_+)\Gamma^a_\mu(q_+,q-k)\nonumber\\
	&&\hspace{1.17cm}\times S(q-k)\Gamma^b_\nu(q_-,q-k)S(q_-)\bigg]\nonumber\\
	&&\hspace{1.17cm}+\textrm{tr}\bigg[
	\Gamma_\eta(P,q)S(q_+)\Gamma^b_\nu(q_+,q+k)\nonumber\\
	&&\hspace{1.17cm}\times S(q+k)\Gamma^a_\mu(q_-,q+k)S(q_-)\bigg]
\Bigg\}\;,\nonumber\\
\end{eqnarray}
where the trace is over Dirac and colour indices, and
the second term accounts for the crossed gluon-exchange.
Solution of this equation requires knowledge of the pseudoscalar 
Bethe-Salpeter amplitude (BSA) for the
$\Gamma_\eta(P,k)$,
the full quark propagator $S(k)$, and the quark-gluon vertex
$\Gamma^a_\mu(k,q)$.

\section{Calculational Inputs\label{Sec:inputs}}
To calculate the quantities required for computation of the diamond
diagram, we employ the Dyson-Schwinger approach to QCD,
see~\cite{Roberts:1994dr,Alkofer:2000wg,Maris:2003vk,Fischer:2006ub} for
reviews. These comprise of an infinite tower of coupled integral
equations that interrelate the fundamental Green's functions of the
theory. Despite analytic studies being possible in certain
kinematical
situations~\cite{Lerche:2002ep,Alkofer:2004it,Fischer:2006vf,Alkofer:2006gz,Alkofer:2008jy},
in principle one must employ a truncation scheme and perform the
calculations numerically. In the present context it is, however, important 
to mention that anomalous
diagrams are represented exactly if the underlying symmetries are respected, see
{\it e.g.\/} refs.~\cite{Frank:1994gc,Alkofer:1995jx}.

\subsection{Yang-Mills Propagators}
\begin{figure}[b]
\includegraphics*[width=0.99\columnwidth]{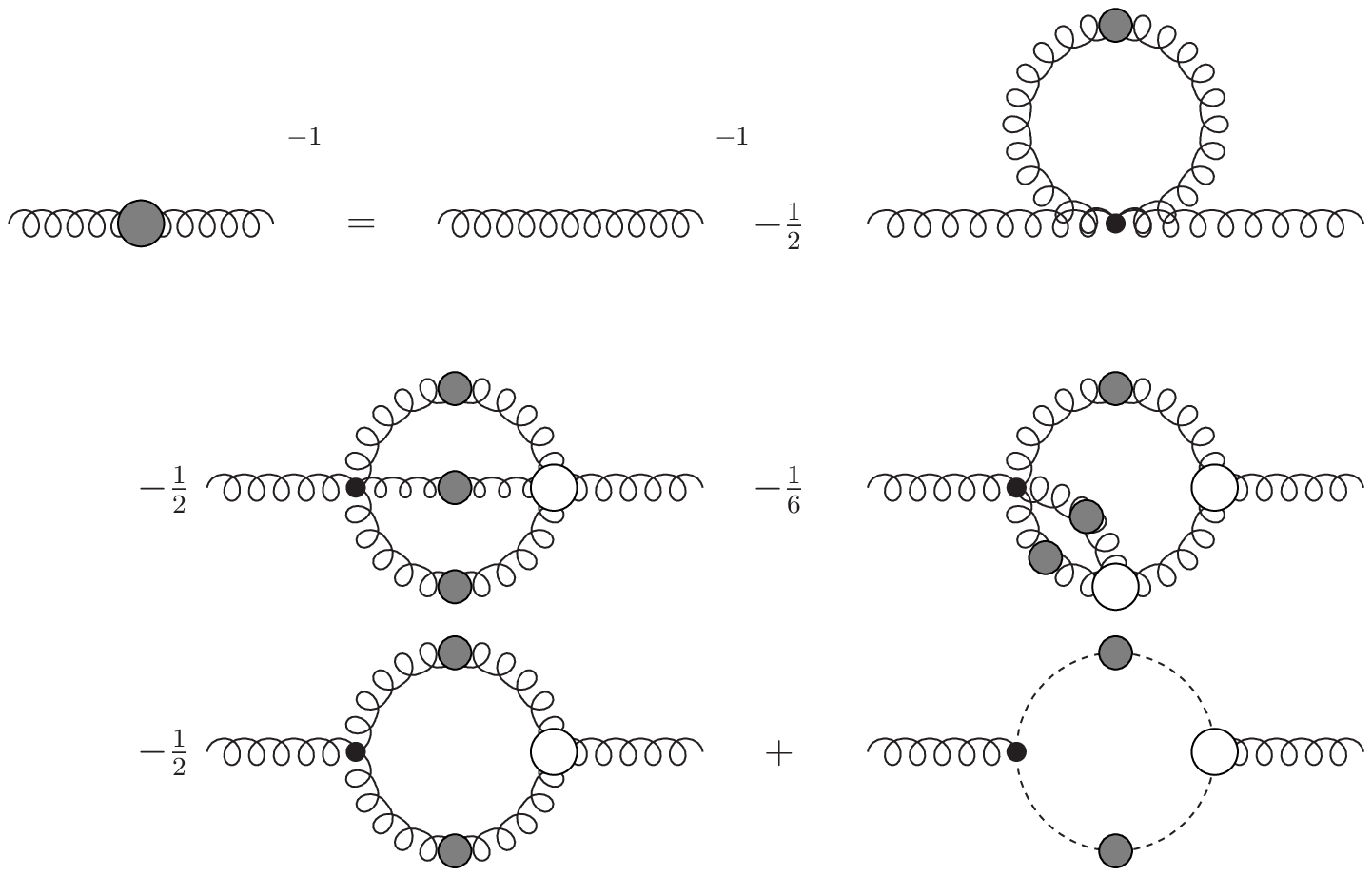}\\[2mm]
\includegraphics*[width=0.99\columnwidth]{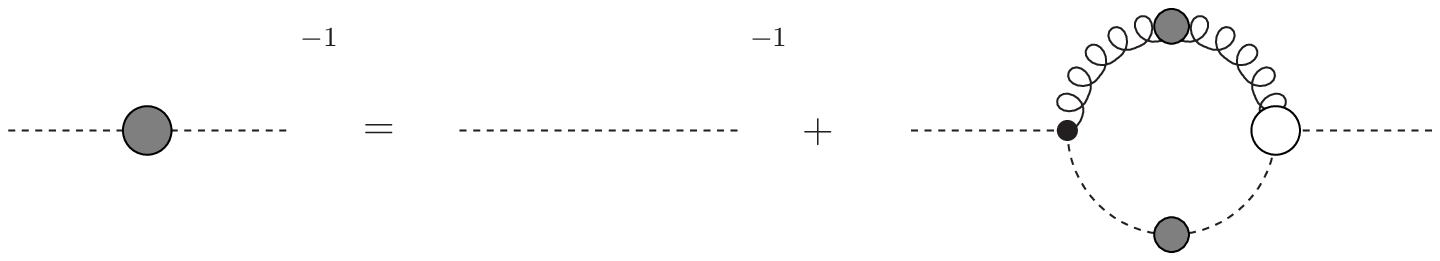}
\caption{The gluon and ghost Dyson-Schwinger equation. Filled blobs indicate
dressed propagators and vertices.}
\label{fig:ghostgluonDSE}
\end{figure}

The pure Yang-Mills (YM) sector of QCD is comprised of both ghosts and
gluons, whose propagators in Euclidean space are written:
\begin{equation}
D^G(p^2) =-\frac{G(p^2)}{p^2}\;,\quad D^{ab}_{\mu\nu}=\delta^{ab}\left(
\delta_{\mu\nu}-\frac{p_\mu p_\nu}{p^2}
\right)\frac{Z(p^2)}{p^2}\;.
\label{eqn:ghostglue}
\end{equation}
where $G(p^2)$ and $Z(p^2)$ are the ghost and gluon dressing functions
respectively. These satisfy the Dyson-Schwinger equations shown
pictorially in fig.~\ref{fig:ghostgluonDSE}, and have been studied in
Landau gauge by employing
truncations at the level of the vertices. It is found that in the infrared
these can be described by power
laws~\cite{von Smekal:1997is,Lerche:2002ep,Zwanziger:2001kw}:
\begin{equation}
G(p^2)\sim \left( p^2 \right)^{-\kappa}\;, \qquad Z(p^2)\sim
(p^2)^{2\kappa}\;,
\label{eqn:irpower}
\end{equation}
with $\kappa$ a positive constant, giving rise to an infrared
diverging ghost propagator and vanishing gluon. Moreover, with a bare
ghost-gluon vertex one finds $\kappa=\left(93-\sqrt{1201}\right)/98\simeq0.595$, and so we see IR
dominance of the ghost in Landau gauge. This fact
greatly affects our calculation of (\ref{eqn:amplitude}) since we
now entirely depend upon the dressing of the quark-gluon vertex to yield
the sufficient infrared enhancement that will lead to non-vanishing
$\Pi(P^2\rightarrow 0)$.

From the ghost-gluon vertex we may define the running coupling:
\begin{equation}
\alpha(p^2)=\alpha_\mu G^2(p^2)Z(p^2)\;,
\label{eqn:running}
\end{equation}
which may be parameterised such that the numerical results
for Euclidean scales are accurately reproduced \cite{Alkofer:2003jj}:
\begin{eqnarray}
\alpha_{\rm fit}(p^2) &=& \frac{\alpha_s(0)}{1+p^2/\Lambda^2_{\rm QCD}}
+ \frac{4 \pi}{\beta_0} \frac{p^2}{\Lambda^2_{\rm QCD}+p^2}
\nonumber\\[-2mm] \label{eqn:alpha}\\[-2mm]
&&\times\left(\frac{1}{\ln(p^2/\Lambda^2_{\rm QCD})}
- \frac{1}{p^2/\Lambda_{\rm QCD}^2 -1}\right)\;. \nonumber
\end{eqnarray}
Here $\beta_0=(11N_c-2N_f)/3$, and $\alpha_S(0)$ is the fixed point in the
infrared, calculated to be $8.915/N_c$ for our choice of $\kappa$.
Similarly, the gluon dressing function
may be fitted~\cite{Alkofer:2003jj}:
\begin{equation}
 Z\left(k^2\right) = \left(\frac{k^2}{k^2+\Lambda^2_{\rm QCD}}\right)^{2\kappa}
                  \left(\frac{\alpha_{\rm
			fit}\left(k^2\right)}{\alpha_\mu}\right)^{-\gamma}\;,
 \label{eqn:gluonfit}
\end{equation}
such that no further singularities are introduced. Here, 
$\gamma = (-13 N_c + 4 N_f)/(22 N_c - 4 N_f)$
is the one-loop value for
the anomalous dimension of the gluon propagator, and $\alpha_\mu=0.2$ at
the renormalisation scale $\mu^2=170$ GeV$^2$. We use $\Lambda^2_{\rm
QCD}=0.5$ GeV$^2$ similar to the scale obtained in
ref.~\cite{Alkofer:2003jj}.

\subsection{The Gap Equation}
\begin{figure}[b]
\includegraphics*[width=0.99\columnwidth]{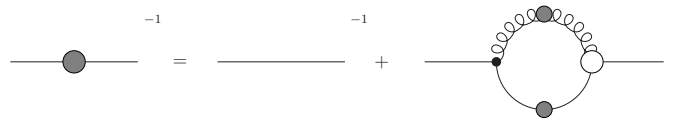}
\caption{The quark Dyson-Schwinger equation. Filled blobs indicate
dressed propagators and vertices.}
\label{fig:quarkDSE}
\end{figure}

A far simpler system to that of the Yang-Mills sector, though one of
equal if not greater importance, is that for the quark propagator, shown in
fig.~\ref{fig:quarkDSE}. This may be written explicitly: 
\begin{eqnarray}\label{eqn:quarksde}	
	&&S^{-1}(p) =  Z_2\left[ S^{(0)}(p) \right]^{-1}\\
	&&~\,\,\,-C_F \frac{\tilde{Z}_1\,Z_2}{\tilde{Z}_3}
	\frac{g^2}{\left( 2\pi \right)^4}\int d^4k
	\gamma_\mu S(k)\Gamma_\nu(k,p)D_{\mu\nu}(p-k)\;. \nonumber
\end{eqnarray}
As is evident, this is a non-linear equation dependent upon the form of
the quark propagator itself, the gluon propagator and a higher order
Green's function, the quark-gluon vertex $\Gamma_\nu(k,p)$. 
The quark propagator consists of a Dirac-odd and Dirac-even part, and so
is completely described by two momentum dependent functions:
\begin{equation}
S(p)=\frac{i\pslash+M(p^2)}{p^2+M^2(p^2)}Z_f(p^2)\;,
\label{eqn:quarkprop}
\end{equation}
with $M(p^2)$ the mass function and $Z_f(p^2)$ the wave-function
renormalisation, which are both assumed to be non-singular in the
infrared. This behaviour is indeed what is seen in numerical 
calculations \cite{Fischer:2003rp,Alkofer:2008tt}.
In the following we consider the quarks in the chiral limit, 
i.e. $[S^{(0)}(p)]^{-1}=i\pslash$. Then $M(p^2)$ only contains 
effects from the dynamical breaking of chiral symmetry and
in the $\eta-\eta'$-mixing we only have the contribution $m_A$ 
from the topological susceptibility (\ref{WV}).

\subsection{Quark-Gluon Vertex}

Earlier we stated that in light of the vanishing
nature of the gluon propagator in the deep IR, it is the behaviour of
the quark-gluon vertex there that is crucial.  

\subsubsection{Vertex for all scales vanishing}
From the IR analysis of~\cite{Alkofer:2006gz,Alkofer:2008tt} we know 
that the quark-gluon vertex, $\Gamma^a_\mu = \frac{\lambda^a}{2} \Gamma_\mu$, 
should have the behaviour:
\begin{equation}
\Gamma\sim \left(\bar{p}^2\right)^{-\kappa-1/2}\;,
\label{eqn:qginfrared}
\end{equation}
when the sum of all incoming momenta, $\bar{p}$, are vanishing in the infrared; 
that is, it is strongly infrared enhanced. They find that all twelve
possible structures of the quark-gluon vertex may contribute in the
infrared, with vector and scalar pieces playing a dominant r\^ole. Thus
one may parameterise $\Gamma_\nu$ by:
\begin{eqnarray}
\Gamma_\nu\left(p,q\right) &=& 
L_1\left(p,\;q\right)\gamma_\nu
			-i\;L_3\left(p,\;q\right) (p+q)_\nu
\label{vertexparam}
\end{eqnarray}
with $p$ and $q$ the incoming and outgoing quark momenta respectively,
and $k=p-q$ the outgoing gluon momentum. 

We wish to find a non-vanishing $\Pi(P^2\rightarrow 0)$, 
and thus look at the quark-triangle $G_{\mu\nu}^{ab}$ in the limit of 
the incident gluon momenta vanishing, i.e. $k,P \ll \Lambda_{\rm QCD}$. 
We note that we may write:
\begin{equation}
G_{\mu\nu}^{ab} = \frac{1}{2}\delta^{ab}\epsilon_{\mu\nu\alpha\beta}P^\alpha
k^\beta\; I\left(P^2,k^2,k\cdot P\right)\;,
\label{eqn:qrkI}
\end{equation}
where $I$ represents the internal loop integral of the quarks, obtained
by separating out the tensor structure. We then evaluate the behaviour
of $I$ numerically, using the power laws (\ref{eqn:irpower}) and 
(\ref{eqn:qginfrared}) for the gluon propagator and the quark-gluon 
vertex, respectively, and an infrared finite quark propagator and 
Bethe-Salpeter amplitude. We then find that the dominant infrared behaviour 
of the triangle is given by:
\begin{equation}
I(P^2,0,0) \sim \left(P^2\right)^{-2\kappa+1}\;.
\label{eqn:Ipower}
\end{equation}
On dimensional grounds this behaviour is insufficient to generate 
$\Pi\ne 0$ in the limit $P^2\rightarrow 0$. We thus conclude that the 
uniform scaling behaviour (\ref{eqn:qginfrared}) of the quark-gluon
vertex is not sufficient to account for the Kogut-Susskind mechanism.  

\subsubsection{Soft Singularity}
However, there is not just one momentum configuration giving rise to
a singular structure in the vertex. There also exists a soft collinear-like
divergence dependent only upon the external gluon momentum $k^2$:
\begin{equation}
\Gamma\sim \left( k^2 \right)^{-\kappa-1/2}
\label{eqn:softdivergence}
\end{equation}
This additional divergence 
has been identified from the Dyson-Schwinger equation for the quark-gluon
vertex in ref.~\cite{Alkofer:2008tt}. Its appearance in the quark-antiquark
scattering kernel for heavy quarks leads to a linearly rising potential and
consequently to quark-confinement. Since the derivation of eq.~(\ref{eqn:softdivergence})
is somewhat involved we do not wish to repeat it here but instead refer 
the interested reader directly to ref.~\cite{Alkofer:2008tt}.
 
For the integrand $I$ at small momenta this soft divergence leads to
\begin{equation}
I(P^2,0,0) \sim \left(P^2\right)^{-2\kappa-1}\;,
\label{eqn:Ipowersoft}
\end{equation}
a behaviour which we have obtained both by analytical power counting 
and also numerically. To gauge the impact of this result on our 
evaluation of $\Pi(P^2\rightarrow 0)$, we insert (\ref{eqn:qrkI}) 
into (\ref{eqn:amplitude}) to
obtain:
\begin{equation}
\Pi(P^2) = 64 \int \widetilde{dk}\; \left(k\cdot P\right)^2
\frac{Z(k_+^2)}{k_+^2}\frac{Z(k_-^2)}{k_-^2} I^2\;.
\label{eqn:gwithI}
\end{equation}
A simple check of the dimensions indicates that a non-zero contribution
to the anomalous mass is obtained. Only the magnitude of this needs
now be determined. To this end, we must still determine the Bethe-Salpeter 
amplitude for the $\eta$. 

\subsection{The Bethe-Salpeter equation}
An essential input to the diamond diagram is the Bethe-Salpeter
amplitude for the $\eta$. This can be obtained by solving the
pseudoscalar Bethe-Salpeter equation (BSE), which describes 
$\qqbar$ bound-states:
\begin{equation}
\Gamma_{tu}(p;P)=\int\widetilde{dk}
	K^{YM}_{tu;rs}(p,k;P)\left[S(k_+)\Gamma(k;P)S(k_-)\right]_{sr}\;,
\label{eqn:bseps}
\end{equation}
where $K$ represents the (unknown) quark-antiquark scattering kernel,
$k_+=k+\eta P$, $k_-=k+(\eta-1)P$, $\eta$ is the momentum partitioning
of the quarks. The meson has total momentum $P$, taken in the rest
frame, with $P^2=-m^2$ and $m$ the mass of the bound-state.

Fundamental to any study of pseudoscalar
mesons is the axial-vector Ward-Takahashi identity (AV-WTI):
\begin{eqnarray}
P_\mu \Gamma_{5\mu}^a(k;P)&=& S^{-1}(k_+)\frac{1}{2}\lambda^a_f i\gamma_5
+\frac{1}{2}\lambda^a_f i\gamma_5 S^{-1}(k_-)\nonumber\\[-2mm]
&&\label{eqn:axialvec}\\[-2mm]
&&-M_\zeta i\Gamma^a_5(k;P)-i\Gamma_5^a(k;P)M_\zeta\;.\nonumber
\end{eqnarray}
Since the left-hand side involves the BSE, whilst the right-hand side
involves the quark SDE, a non-trivial relationship between the kernels
is established. Thus one must treat the Bethe-Salpeter kernel $K$ and the
quark-gluon vertex in such a way that the AV-WTI is satisfied, otherwise 
a massless pion will not be found in the chiral limit.  A generalisation
of this identity to the case of the flavour singlet channel has been
given in ref.~\cite{Bhagwat:2007ha}. A suitable
truncation scheme used frequently \cite{Maris:2003vk} is the 
rainbow-ladder approximation which we also employ here. 

\subsection{Phenomenologically qualitative model}
In rainbow-ladder approximation the quark-gluon vertex is restricted
to its $\gamma_\nu$ part. We therefore simplify the parametrisation  
(\ref{vertexparam}) further and write:
\begin{equation}
\Gamma_\nu(k,p) =
Z_{1F}(\mu^2,\Lambda^2) L_1(k^2,\mu^2) \gamma_\nu\;.
\end{equation}
Here the vertex
renormalisation constant $Z_{1F}$ provides multiplicative 
renormalisability of the vertex and also of the resulting quark SDE 
and pseudoscalar BSE. The vertex dressing $L_1$ is parameterised by:
\begin{eqnarray}
L_1\left(z\right) &=& \left(\frac{z}{z+
d_2}\right)^{-1/2-\kappa}\Bigg( \frac{ d_1}{1+z/{ d_2}}+
\frac{ z d_3}{d_2^2+\left(z- d_2\right)^2}
\nonumber\\[-1.2mm]
&+&\frac{z}{ { d_2}+z}\left[\frac{4\pi}{\beta_0\alpha_\mu}
\left(\frac{1}{\log\left(z/{d_2}\right)}-\frac{d_2}{z-d_2}\right)\right]^{-2\delta}
	    \Bigg)\nonumber \\
\label{eqn:l1param}
\end{eqnarray}
with $z=k^2$ the gluon momentum and $d_1$ sets the strength of the interaction
in the infrared. The scale $d_2$ is set to 
$d_2=(0.5{\rm GeV})^2 \approx \Lambda_{\rm QCD}^2$, 
i.e. similar to the scale found in the Yang-Mills-sector.
The parameter $d_3$ determines the size of a term added to give additional 
integrated strength in the intermediate momentum regions. 

This construction is necessary to produce meaningful results: Since the
AV-WTI forces us at present to only consider the $\gamma_\mu$-part of the
quark-gluon interaction we necessarily miss some interaction strength 
in the intermediate momentum region. This cannot be compensated for by an 
increase of $d_1$ as discussed in the following.

\begin{figure}[h]
\includegraphics*[width=0.99\columnwidth]{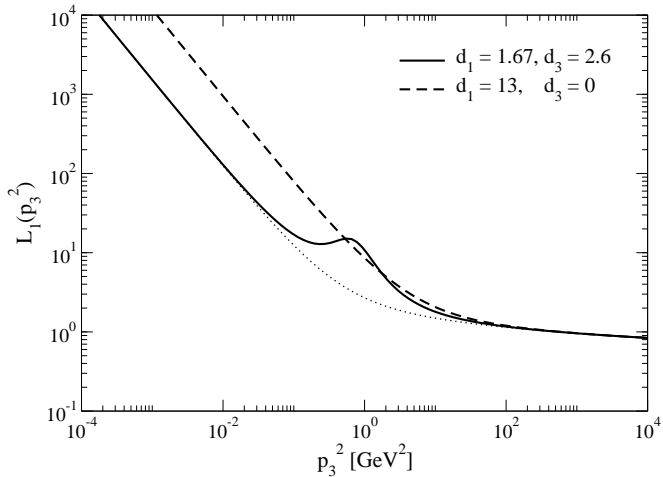}
\caption{Our model dressing for the $\gamma^\mu$ part of the vertex.
See the text for a description of each.}
\label{fig:model}
\end{figure}

In fig.~\ref{fig:model} we plot two examples of $L_1$. The parameters of the dashed line have been fitted such that the physical masses of the $\pi$
and the $\rho$ are reproduced, employing $d_1$ and $d_2$ only. Here we need a 
large value of $d_1=13$, which leads to an extremely strong infrared singularity; 
employing such a form in the diamond diagram would yield results for the topological mass $m_A^2$ which are orders
of magnitudes too large to account for the experimental $\eta$ and
$\eta'$-masses. To account for this, we introduce extra interaction strength in
the intermediate momentum region parameterised by $d_3$. This additional
bump generates correct $\pi$ and $\rho$-masses whilst allowing for a reduction of the infrared scale 
$d_1$ to a realistic order. Note, that this particular shape and
strength of interaction is also suggested by the results for the running coupling of
the quark-gluon vertex, found in~\cite{Alkofer:2008tt}.

Once our parameters are fixed, we obtain solutions 
for the Bethe-Salpeter amplitude in the chiral limit, which are then employed
in the calculation of the diamond diagram.

\section{Results\label{Sec:results}}
\begin{table*}[t]
\centering
\begin{tabular}{ccc|cc|cc|ccc}
$d_1$  & $d_2$   & $d_3$   & $m_\pi$ & $m_\rho$ & $m_A^2$ &  $\chi^2$    & $\theta$& $m_\eta$ & $m_{\eta'}$\\
GeV$^2$& GeV$^2$ & GeV$^2$ &  MeV    &   MeV    & GeV$^2$ & (MeV)$^4$ &         &   MeV    & MeV        \\
\hline\hline
$1.41$ & $0.5$   & $2.6$   & $135$   &  $735$   & $0.30$  & $144$      & $-35.3$ &   $412$  & $790$	  \\
$1.55$ & $0.5$   & $2.6$   & $135$   &  $741$   & $0.48$  & $162$      & $-29.1$ &   $450$  & $840$	  \\
$1.67$ & $0.5$   & $2.6$   & $135$   &  $747$   & $0.56$  & $169$      & $-23.2$ &   $479$  & $906$	  \\
\end{tabular}
\caption{Numerical results of the calculation of the diamond diagram for
a range of parameters. The $\pi$ and $\rho$ phenomenology is only
marginally 
affected by these small changes in infrared strength, yet note the
strong effect on the calculated anomalous mass.\label{tab:results}}
\end{table*}

Na\"ively, the diamond diagram consists of a twelve dimensional integral
which can be reduced to eight dimensions by symmetry considerations.
In the limit $P^2\rightarrow0$ the integrand simplifies further and the 
problem may be reduced to a five dimensional integral. We solved this
integral both, by Monte-Carlo techniques and using adaptive quadrature
routines and found the latter method to be superior. Sufficient accuracy
of the order of one percent is obtained within a few minutes of CPU-time.

In table \ref{tab:results} we give our results for the calculation of
the diamond diagram shown in fig.~\ref{fig:diamond}, employing the model
of the quark-gluon interaction discussed above. Within the limitations of 
the model we may not expect quantitatively correct results. Nevertheless
we obtain reasonable values for the masses and the mixing angle at least
for the last parameter set considered. For this set, we find a
topological susceptibility,
\begin{equation}
	\chi^2 = \left( 169\; \textrm{MeV}\right)^4\;,
\end{equation}
in qualitative agreement with lattice results~\cite{DelDebbio:2004ns}.
Qualitatively we do see that it is
precisely the infrared collinear behaviour of the quark-gluon vertex that 
is necessary to obtain a non-zero contribution to the topological
susceptibility.

A quantitative study would require a self-consistent solution for the
quark-gluon vertex employing fewer approximations. Then, at least in the
$P^2\rightarrow0$ limit, one need not solve the pseudoscalar BSE since the
leading term $\gamma_5 B(q^2)/f_\pi$ in chiral limit is known from the
quark DSE. We would then need to consider higher-order diagrams
corresponding to the diamond diagram, with $n$-gluon exchange taking
place in the s-channel. This ladder re-summation of gluons is often 
represented by a single meson exchange. This picture has been explored in 
ref.~\cite{Lakhina:2007vm} as a means of generating the topological mass
of the $\eta'$.
However, since the actual mechanism of anomalous $U_A(1)$ symmetry
breaking may not be explained in terms of gauge independent particle
exchange, such as a pion, this picture cannot be entirely adequate.

\section{Discussion of topological effects and infrared singularities \label{topo}}

As we have seen the infrared divergence of the quark-gluon vertex plays an
important r\^ole in a confinement-based explanation of the  $\eta^\prime$ mass 
and thus the $U_A(1)$ anomaly. On the other hand, it is evident that the
$\eta^\prime$ mass is linked to the topological susceptibility 
via the Witten-Veneziano formula (\ref{WV})
and thus topologically non-trivial gluon field configurations. How can such a dichotomy
be resolved?

In a first step to a potential solution of this puzzle we note that it
is not only
instantons that are capable of providing a non-vanishing topological susceptibility.
As a matter of fact, there is evidence 
\cite{Ilgenfritz:2005ga,Ilgenfritz:2005um,Gattringer:2002wh} that on the
lattice,
field configurations with non-trivial holonomy have been misidentified
as instantons. They are more like Kraan--van-Baal--Lee--Lu 
calorons \cite{Kraan:1998pm,Lee:1998bb} than instantons. And, amongst these and
other topologically non-trivial gluon field configurations, also center vortices
(see {\it e.g.\/} the reviews \cite{Greensite:2007zz}) provide a non-vanishing
topological susceptibility \cite{Engelhardt:2000wc,Bertle:2001xd}. At least, for
these latter field configurations we know that they provide an explanation for
quark confinement \cite{Greensite:2007zz,Alkofer:2006fu}, and that in most
gauges, including the Landau gauge, a substantial amount of these
configurations live on the Gribov horizon. 
The latter property is responsible that, when removing center vortices  
from a lattice ensemble, not only the string tension vanishes 
but also the Landau gauge ghost propagator becomes
infrared finite \cite{Gattnar:2004bf}.

In the confinement based scenario described within this letter, the infrared
divergence of the quark-gluon vertex is the cause for the $\eta^\prime$ mass. It
is driven by the infrared divergence of the three-gluon gluon vertex and thus
eventually by the infrared behaviour of the ghost propagator
\cite{Alkofer:2006gz}. In Landau gauge (as well as in some other gauges) this
infrared divergence is, according to the Gribov-Zwanziger scenario, assumed to
be caused by the dominance of field configurations on the Gribov horizon (or,
more precisely, on the non-vanishing overlap of the boundary of the fundamental
modular region with the Gribov horizon). At least, some of these field
configurations, as {\it e.g.\/}  the center vortices, are topologically
non-trivial. If the Gribov-Zwanziger scenario is correct they provide
confinement. But they may also provide a  non-vanishing topological
susceptibility and thus the $U_A(1)$ anomaly, especially the $\eta^\prime$ mass.
When using, instead of a lattice Monte-Carlo calculation, a functional
approach (as Dyson-Schwinger or Renormalization Group Equations)
topological effects are encoded in the infrared behaviour of the Green
functions. In Landau gauge
it is the chain of infrared divergences of the ghost propagator, the three-gluon
and the quark-gluon vertex which seems to be responsible for both, confinement and the
$U_A(1)$ anomaly. This is, however, only a reflection of the fact that certain
types of gluon field configurations cause both, confinement and the
$U_A(1)$ anomaly.

If one is willing to accept this scenario a further conclusion is evident: the
$U_A(1)$-relevant, topologically non-trivial gluon field configurations are
intimately related to confinement, and thus are very likely not the
instantons discovered already more than thirty years ago \cite{Belavin:1975fg}.
In addition, we know at least one type of gluon field configuration
satisfying the requirements of this scenario, namely center vortices.
However, at this point the relevance of other configurations to confinement and
the $U_A(1)$ anomaly cannot be excluded. 

\section{Summary \label{Sec:sum}}
We investigated the Kogut-Susskind mechanism for the resolution of the
$U_A(1)$ problem of QCD, employing known results for the ghost and gluon
propagators from Landau gauge studies of the Yang-Mills sector, together
with recent results for the quark-gluon vertex. Taking the qualitative 
features of this study, we modelled
the vertex-dressing such that an infrared soft-singularity was present,
whilst at the same time matching meson observables for the $\pi$ and the
$\rho$. The presence of a collinear singularity \cite{Alkofer:2008tt}, 
dominant in the gluon momentum, proves essential to the generation of a 
non-zero topological susceptibility in the chiral limit,
as required by the Witten-Veneziano mechanism. Our study illustrates 
the effects of topologically non-trivial field configurations on the level 
of correlation functions and therefore provides a 
qualitatively feasible mechanism for the resolution of the mass of the 
$\eta'$ in a functional approach to QCD.

\section*{Acknowledgements}
We are grateful to Felipe Llanes-Estrada, Craig Roberts, Kai Schwenzer and
Lorenz von Smekal for fruitful discussions on this problem. 
RW thanks the hospitality of Graz where much of
this work was undertaken.
RA and RW have been supported  by the Deutsche Forschungsgemeinschaft
(DFG) under Grant No.\ Al279/5-2,
CF and RW have been 
supported by the Helmholtz-University Young Investigator Grant VH-NG-332.

\end{document}